\documentstyle[11pt]{article}
\newfont{\gothic}{eufm10 scaled\magstep 1}
\def\lp{\stackrel{\leftarrow}{\partial}}
\def\rp{\stackrel{\rightarrow}{\partial}}

\def\bigstar{\;\bigcirc\kern-0.9em\star\;}  
\begin{document}
\begin{flushright} 
DTP/97/61\\ MIAMI-TH-97-3 \\ ANL-HEP-PR-97-93    
\end{flushright}
{\Large
\centerline{FEATURES OF TIME-INDEPENDENT WIGNER FUNCTIONS}
\centerline{Thomas Curtright$^{\S}$, David Fairlie$^\natural$, 
and Cosmas Zachos$^{\sharp}$} }

$^{\S}$ Department of Physics, University of Miami,
Box 248046, Coral Gables, Florida 33124, USA\\
\phantom{.} \qquad\qquad{\sl curtright@phyvax.ir.Miami.edu}  

$^{\natural}$ Department of Mathematical Sciences, University of Durham, 
Durham, DH1 3LE, UK  \\  
\phantom{.} \qquad\qquad{\sl David.Fairlie@durham.ac.uk}

$^{\sharp}$ High Energy Physics Division,
Argonne National Laboratory, Argonne, IL 60439-4815, USA \\
\phantom{.} \qquad\qquad{\sl zachos@hep.anl.gov}      

\begin{abstract}
The Wigner phase-space distribution function provides the basis for 
Moyal's deformation quantization alternative to the more conventional 
Hilbert space and path integral quantizations. 
General features of time-independent Wigner functions are explored here, 
including the functional (``star") eigenvalue equations they satisfy;  
their projective orthogonality spectral properties;
their Darboux (``supersymmetric") isospectral potential recursions; 
and their canonical transformations. These features are illustrated 
explicitly through simple solvable potentials: the harmonic 
oscillator, the linear potential, the P\"{o}schl-Teller potential, 
and the Liouville potential.

\end {abstract}

\vskip 0.5cm

\section{Introduction}

Wigner functions have been receiving increasing attention in quantum 
optics, dynamical systems, and the algebraic structures of M-theory 
\cite{reviews}. They were invented by Wigner and Szilard \cite{wigner}, 
and serve as a phase-space distribution alternative to the density matrix,
to whose matrix elements they are related by Fourier transformation. 
The diagonal, hence real, time-independent pure-state Wigner function 
$f(x,p)$ corresponding to the eigenfunction $\psi$ of 
${\bf H}\psi=E\psi$, is 
\begin{equation}
f(x,p)={1\over 2\pi}\int\! dy~\psi^* (x-{\hbar\over2} y )~e^{-iyp} 
\psi(x+{\hbar\over2} y).    \label{orururumo}               
\end{equation}
These functions are not quite probability distribution functions, as they 
are not necessarily positive---illustrated below. However, upon integration 
over $p$ or $x$, they  yield bona-fide positive probability distributions, 
in $x$ or $p$  respectively. 

Wigner functions underlie Moyal's formulation of Quantum Mechanics 
\cite{moyal}, through the unique \cite{vey,bayen} one-parameter 
($\hbar$) associative deformation of the Poisson-Bracket structure of 
classical mechanics. Expectation values can be computed on the basis 
of phase-space c-number functions: given an operator ${\bf A(x,p)}$, 
the corresponding phase-space function $A(x,p)$ obtained by 
${\bf p}\mapsto p,~~{\bf x}\mapsto x$ 
yields that operator's expectation value through
\begin{equation}
\langle {\bf A} \rangle =\int\! dx dp~ f(x,p) A(x,p) ~,
\end{equation}
assuming the usual normalization, $ \int dx dp f(x,p)=1$, and further 
assuming Weyl ordering, as addressed by Moyal, who took matrix elements 
of all such operators: 
\begin{equation}
{\bf A(x,p)}=\frac{1}{(2\pi)^2}\int d\tau d\sigma 
dx dp ~A(x,p) \exp (i\tau ({\bf p}-p)+i\sigma ({\bf x}-x))~.
\end{equation}

Wigner functions are c-numbers, but they compose with each other 
nonlocally. Properties of these compositions were explored in, e.g., 
\cite{baker,dbfcam}, and were codified in an elegant system in 
\cite{bayen}: to parallel operator multiplication, the Wigner functions 
compose with each other through the {\em associative} star product 
\begin{equation} 
\star \equiv e^{{i\hbar \over 2} (\lp_x \rp_p- \lp_p \rp_x )} .
\end{equation}
Recalling the action of a translation operator 
$\exp(a \partial_x)~h(x)=h(x+a)$,
it is evident that the $\star$-product induces simple ``Bopp"-shifts:
\begin{equation}             
f(x,p) \star g(x,p) = 
f(x,p-{i\hbar\over 2} \rp_x)~ g(x,p+{i\hbar\over 2} \lp_x) = 
f(x+{i\hbar\over 2} \rp_p ,p-{i\hbar\over 2} \rp_x)~ g(x,p),
\end{equation}
etc., where $\lp$ and $\rp$ here act on the arguments of $f$ and $g$, 
respectively. This intricate convolution samples the Wigner function 
over the entire phase space, and thus provides an alternative 
to operator multiplication in Hilbert space.

Antisymmetrizing and symmetrizing the star product, yields the Moyal (Sine) 
Bracket \cite{moyal},
\begin{equation} 
\{\{f,g\}\} \equiv {f\star g - g\star f \over 2i}, 
\end{equation}
and Baker's \cite{baker} Cosine Bracket 
\begin{equation}
(( f,g))  \equiv {f\star g + g\star f\over 2 }, 
\end{equation}
respectively. Note \cite{dbfcam,hansen} that 
\begin{equation}
\int\! dp dx ~ f\star g = \int\! dp dx ~ fg~ \label{Ndjambi}.
\end{equation}
Further note the Wigner distribution has a $\star$-factorizable integrand,
\begin{equation}
f(x,-2p)= {1\over 2\pi} \int\! dy ~\left (\psi^* (x)~  e^{iyp}\right)\star 
\left(\psi(x)~  e^{iyp}\right) .  
\end{equation}

In general, systematic specification of time-dependent Wigner functions is 
predicated on the eigenvalue spectrum of the time-independent problem. 
For pure-state static distributions, Wigner and, more explicitly, Moyal, 
showed that 
\begin{equation}             
\{\{ H(x,p),f(x,p)\}\} = 0, \label{Karunga}
\end{equation}
i.e.~ $H$ and $f$ $\star $-commute. 
However, there is a more powerful functional equation, the ``star-genvalue"   
equation, which holds for the
time-independent pure state Wigner functions (Lemma 1), and amounts to a
complete characterization of them (Lemma 2). 

We will explore the features of this $\star$-genvalue equation, and 
illustrate its utility on a number of solvable potentials, including both 
the  harmonic oscillator and the linear one.
The $\star$-multiplications of Wigner functions will be seen to
parallel Hilbert space operations in marked detail.
The P\"{o}schl-Teller
potential will reveal how the hierarchy of factorizable Hamiltonians
familiar from supersymmetric quantum mechanics finds its full analog in
$\star$-space. We determine the Wigner function's transformation
properties under (phase-space volume-preserving) canonical transformations,
which we finally elaborate in the context of the Liouville potential. 

\section{The $\star$-genvalue Equation}
\noindent {\bf Lemma 1}: {\em Static, pure-state Wigner functions obey the 
$\star$-genvalue equation},
\begin{equation}
H(x,p)\star f(x,p) = E ~f(x,p)~.
\end{equation}

\noindent Without essential loss of generality, consider 
$H(x,p)= p^2/2m  + V(x)$. 
\begin{eqnarray}             
H(x,p)\star f(x,p) &=& {1\over 2\pi} \left( (p-i{\hbar\over2} \rp_x)^2/2m 
+V(x) \right) \int\! dy~ e^{-iy(p+i{\hbar\over2} \lp_x)}
\psi^*(x-{\hbar\over2} y) ~\psi(x+{\hbar\over2} y)   \\ 
&=& {1\over 2\pi} \int\! dy~ \left( (p-i{\hbar\over2} \rp_x)^2/2m 
+V(x+{\hbar\over2}y)\right)   e^{-iyp}\psi^*(x-{\hbar\over2} y)~
\psi(x+{\hbar\over2} y) \nonumber \\
&=& {1\over 2\pi} \int\! dy ~e^{-iyp} 
\left( (i\rp_y +i{\hbar\over2} \rp_x)^2/2m + V(x+{\hbar\over2}y)\right)  
\psi^*(x-{\hbar\over2} y) ~\psi(x+{\hbar\over2} y)  \nonumber 
\\ &=& {1\over 2\pi} \int\! dy ~e^{-iyp} \psi^*(x-{\hbar\over2} y)~
E~ \psi(x+{\hbar\over2} y)= E ~f(x,p),\nonumber 
\end{eqnarray}
since the action of the effective differential operators on 
$\psi^*$ turns out to be null; and, likewise, 
\begin{equation}             
f\star H  ={1\over 2\pi} \int\! dy ~e^{-iyp} 
\left( -(\rp_y -{\hbar\over2} \rp_x)^2/2m + V(x-{\hbar\over2}y)\right)  
\psi^*(x-{\hbar\over2} y) ~\psi(x+{\hbar\over2} y)=E f(x,p).
\end{equation}
Thus, both of the above relations (\ref{Karunga}) and Lemma 1 obtain.
{\rule{.4\baselineskip}{.44\baselineskip}}

This time-independent equation was introduced in ref \cite{dbfcam}, 
such that the expectation of the energy $H(x,p)$ in a pure state 
time-independent Wigner function $f(x,p)$ is given by 
\begin{equation}             
\int\! H(x,p)f(x,p)~dxdp = E\int\! f(x,p)~dxdp. 
\end{equation}
On account of the integration property of the star product, 
(\ref{Ndjambi}), the left hand side of this amounts 
to $\int\! dxdp~ H(x,p)\star f(x,p)$. Implicitly, this equation 
could have been inferred from the Bloch equation of the 
temperature- and time-dependent Wigner function, in the early work 
of \cite{oppenheim}. $\star$-genvalue equations are discussed in 
some depth in the second of refs \cite{bayen}, and in \cite{takahashi}.

By virtue of this equation, Fairlie also derived the general 
$\star$-orthogonality and spectral projection properties of 
static Wigner functions \cite{dbfcam}. His results were later 
formalized in the spectral theory of the second of refs 
\cite{bayen} (e.g.~eqn 4.4). Consider $g$ corresponding to the 
(normalized) eigenfunction $\psi_g$ corresponding to energy $E_g$.  
By Lemma 1, and the associativity of the $\star$-product,
\begin{equation}  
f\star H\star g= E_f ~f\star g =E_g ~f\star g.           
\end{equation}
Then, if $E_g\neq E_f$, this is only satisfied by 
\begin{equation}             
 f\star g= 0   \label{omuhona}.
\end{equation}
N.B. ~the integrated version is familiar from Wigner's paper, 
\begin{equation}             
\int\! dx dp~ f\star g= \int\! dx dp~ fg= 0,
\end{equation}
and demonstrates that all overlapping Wigner functions cannot be everywhere 
positive.
The unintegrated relation introduced by Fairlie appears local, but is, 
of course, highly nonlocal, by virtue of the convolving action of the 
$\star$-product.

Precluding degeneracy, for $f=g$, 
\begin{equation}  
f\star H\star f= E_f ~ f\star f= H\star f\star f ,
\end{equation}
which leads, by virtue of associativity, to the normalization relation 
\cite{baker}, 
\begin{equation}             
f\star f \propto f. \label{Nguarorerue}
\end{equation}
Both relations (\ref{omuhona}) and (\ref{Nguarorerue}) can be checked directly:
\begin{eqnarray}             
&&f(x,p)\star  g(x,p) 
= f(x,p-{i\hbar\over 2} \rp_x)~ g(x,p+{i\hbar\over 2} \lp_x)\\ 
&&={1\over (2\pi)^2} \int\! dy ~\psi_f^* (x-{\hbar\over2} y )~  
\psi_f(x+{\hbar\over2} y)~e^{-iy(p-{i\hbar\over 2} \rp_x)} ~ 
\int\! dY ~e^{-iY(p+{i\hbar\over 2} \lp_x)}~ 
\psi_g^* (x-{\hbar\over2}Y)\psi_g(x+{\hbar\over2} Y) \nonumber \\
&&= {1\over( 2\pi)^2} \int\! dy dY  ~e^{-i(y+Y)p}~ 
\psi_f^* (x-{\hbar\over2} y+{\hbar\over 2}Y )~  
\psi_f(x+{\hbar\over2} y+{\hbar\over 2}Y )~~
\psi_g^* (x-{\hbar\over2} Y-{\hbar\over 2}y  )~  
\psi_g(x+{\hbar\over2} Y -{\hbar\over 2}y ) = \nonumber 
\end{eqnarray}
$$
\left( {1\over 2\pi}\! \int\! \! d(Y+y) ~e^{-i(y+Y)p}~ 
\psi_g^* (x-{\hbar\over2} (Y+y))~  
\psi_f(x+{\hbar\over2}( y+Y) )\right)
\left( {1\over h}\! \int\!\! d({\hbar (Y-y)\over 2}) 
\psi_f^* ({\hbar\over 2}(Y-y) )\psi_g({\hbar\over2}(Y-y))\right).  
$$
The second integral factor is 0 or $1/h$, depending on $f\neq g$ or $f=g$, 
respectively, specifying the normalization $f*f =  f/h$ in 
(\ref{Nguarorerue}). In conclusion,

\noindent {\bf Corollary 1}:\qquad \qquad \qquad \qquad \qquad \qquad 
$f_a\star f_b= \frac{1}{h}~ \delta_{a,b}~ f_a$~. 

These spectral properties are summoned up by their own necessity; 
much of their meaning, nevertheless, resides in their margins: 
For non-normalizable wavefunctions, the above second integral factor may 
diverge, as illustrated below for the linear potential, but the orthogonality
properties still hold. 

Thus, e.g., for an arbitrary function(al) $F(z)$,
\begin{equation} 
F[f\star ] f= F(1/h) f ,
\end{equation}
and, for $\star$-genfunctions of Lemma 1, 
\begin{equation} 
F[H\star ] f= F(E) f .
\end{equation}

Baker's converse construction extends to a full converse of Lemma 1, namely 

\noindent {\bf Lemma 2}: {\em Real solutions of 
$H(x,p)\star f(x,p) = E ~f(x,p)~~ (=f(x,p)\star H(x,p))$ must 
be of the Wigner form,  ~$f=\int\! dy~ e^{-iyp} 
\psi^*(x-{\hbar\over2}y) \psi(x+{\hbar\over2}y) /2\pi$, ~ such 
that ~}${\bf H}\psi=E\psi$.

\noindent As seen above, the pair of $\star $-eigenvalue equations 
dictate, for $f(x,p)=\int\!  dy~ e^{-iyp} \tilde{f}(x,y)$,  
\begin{equation}             
\int\! dy ~e^{-iyp} \left(-{1\over 2m}
(\rp_y \pm{\hbar\over2} \rp_x)^2 + V(x\pm{\hbar\over2}y)-E\right)
\tilde{f}(x,y)=0.
\end{equation}
This constrains $\tilde{f}(x,y)$ to consist of bilinears 
$\psi^*(x-{\hbar\over2} y) ~\psi(x+{\hbar\over2} y)$ of unnormalized 
eigenfunctions $\psi(x)$ corresponding to the same eigenvalue $E$ in the 
Schr\"{o}dinger equation with potential $V$. \qquad 
{\rule{.4\baselineskip}{.44\baselineskip}}

These two Lemmata then amount to the statement that, {\em for real 
functions $f(x,p)$, the Wigner form is equivalent to compliance with the 
$\star$-genvalue equation} (real and imaginary part). 

\section{Example: The Simple Harmonic Oscillator}
The eigenvalue equation of Lemma 1 may be solved directly to 
produce the Wigner functions for specific potentials, without first solving 
the corresponding Schr\"{o}dinger problem (as in, e.g. \cite{bartlett}). 
Following \cite{dbfcam}, for the harmonic oscillator, 
$H=(p^2 +x^2)/2$ (with $\hbar=1$, $m=1$), the resulting equation is 
\begin{equation} 
\left( (x+{i\over 2} \partial_p)^2+(p-{i\over 2} \partial_x)^2 
-2E \right)  f(x,p)=0 .
\end{equation}
By virtue of its imaginary part, $(x\partial_p -p \partial_x)f =0$, 
$f$ is seen to depend on only one variable, 
$z=4H=2(x^2+p^2)$, so the equation reduces to a simple ODE,
\begin{equation}     
\left({z\over 4}-z\partial_z^2 -\partial_z -E\right) f(z) = 0.
\end{equation}
Moreover, setting $f(z)=\exp(-z/2) L(z)$, this yields 
\begin{equation}
\left( z\partial_z^2 + (1-z)\partial_z +E-{1\over 2} \right) L(z) = 0,
\end{equation}
which is the equation satisfied by Laguerre polynomials,  
$L_n=e^z \partial^n (e^{-z} z^n)$, for $n=E-1/2=0,1,2,...$, so that 
the un-normalized eigen-Wigner-functions are 
\begin{equation} 
f_n=e^{-2H}L_n(4H); \qquad \qquad \qquad \qquad 
L_0= 1,\quad  L_1=1-4H,  \quad L_2=16H^2-16 H +2, ...  
\end{equation}
Note the eigenfunctions are not positive definite, and are the only ones 
satisfying the boundary conditions, $f(0)$ finite, and $f(z)\rightarrow 0$, 
as $z\rightarrow\infty$.

In fact, Dirac's Hamiltonian factorization method for 
algebraic solution carries through (cf.~\cite{bayen}) 
intact in $\star$-space. Indeed,
\begin{equation}
H= \frac{1}{2} (x-ip) \star (x+ip) + \frac{1}{2},
\end{equation}
motivating definition of 
\begin{equation}
a\equiv \frac{1}{\sqrt{2}} (x+ip),   \qquad \qquad 
a^{\dagger} \equiv \frac{1}{\sqrt{2}} (x-ip).
\end{equation}
Thus, noting
\begin{equation} 
a \star     a^{\dagger} -a^{\dagger} \star a=1,  
\end{equation}
and also that, by above, 
\begin{equation} 
a \star f_0= \frac{1}{\sqrt{2}} (x+ip) \star e^{-(x^2+p^2)}=0,
\end{equation}
provides a $\star$-Fock vacuum, it is evident that associativity 
of the $\star$-product permits the entire ladder spectrum 
generation to go through as usual. The $\star$-genstates of 
the Hamiltonian, s.t.~$H\star f= f\star H$, are thus
\begin{equation}
f_n\propto (a^{\dagger}\star)^n   ~ f_0 ~(\star a)^n ~.
\end{equation}
These states are real, like the Gaussian ground state, and are thus 
left-right symmetric $\star$-genstates. They 
are also transparently $\star$-orthogonal for different eigenvalues; 
and they project to themselves, as they should, since the Gaussian ground 
state does, $f_0 \star f_0 \propto f_0$. It will be seen below that 
even the generalization of this 
factorization method for isospectral potential pairs 
goes through without difficulty.

\section{Further Example: The Linear Potential} 
For simplicity, take $m=1/2$, $\hbar=1$. Recall (\cite{curtghand}) that 
the problem readily reduces to a free particle:
$H(x,p)=p^{2}+x\mapsto H_{free}=P$ is accomplished by canonically 
transforming through the generating function 
$F(x,X)=-\frac{1}{3}\,X^{3}-xX$. 
The energy eigenfunctions are Airy functions,
\begin{equation}
\psi_{E}(x)=\frac{1}{2\pi }\int_{-\infty }^{+\infty}dX\,
e^{iF(x,X)}\,e^{iEX}=Ai(x-E)\;.
\end{equation}
The $\star$-genvalue equation in this case is 
\begin{equation}
\left( (x+{\frac{i}{2}}\partial _{p})+(p-{\frac{i}{2}}\partial
_{x})^{2}-E\right) f(x,p)=0\;,
\end{equation}
whose imaginary part, $\left( \frac{1}{2}\partial _{p}
-p\partial _{x}\right) f(x,p)=0$, gives $f(x,p)=f(x+p^{2})=f(H)$. 
The real part of the equation is
then an ordinary second order equation, just as in the above harmonic 
oscillator case. Moreover, here the real part of the $\star$-genvalue 
equation is essentially the same as the usual energy eigenvalue equation: 
\begin{equation}
\left( z-{\frac{1}{4}}\partial _{z}^{2}-E\right) f(z)=0\;,
\end{equation}
where $z=x+p^{2}$. Hence, the Wigner function is again an Airy function, like
the above wavefunctions, except that the argument has a different
scale and shift\footnote{This case is similar to the Gaussian wave function, 
i.e.~the harmonic oscillator ground state encountered above, whose Wigner 
function is also a Gaussian, but of different width. S Habib kindly informed
us that this solution is also given in ref \cite{habib}, 
eqn (29).}. 
\begin{equation}
f(x,p)=\frac{2^{2/3}}{2\pi}Ai(2^{2/3}(z-E))
=\frac{2^{2/3}}{2\pi}Ai(2^{2/3}(x+p^{2}-E))
={\frac{1}{(2\pi)^2}}
\int\! dy~ e^{iy(E-x-p^2-y^2/12)}\;.  \label{Herero}
\end{equation}
The Airy functions are
not square integrable, so that the conventional normalization 
$f\star f=\frac{1}{2\pi }\,f$ does not strictly apply. On the other hand, 
the energy eigenfunctions are non-degenerate, and the general 
Corollary 1 projection relations $f_a\star f_b\propto \delta_{a,b} f_a$ 
{\em still} hold for the continuous spectrum: 
\begin{eqnarray}
f_{E1}\star f_{E2} &=&f_{E1}((x+{\frac{i}{2}}\rp _{p})
+p^{2})~f_{E2}((x-{\frac{i}{2}}\lp _{p})+p^{2})  \nonumber \\
&=&{\frac{1}{(2\pi)^{4}}}  \int dy~dY~e^{iy(E1-x-(p-Y/2)^{2}
-y^{2}/12)}~e^{iY(E2-x-(p+y/2)^{2}-Y^{2}/12)} \nonumber \\
&=&{\frac{1}{(2\pi)^{4}}}  \int \!d(y+Y)~e^{i(y+Y)
({\frac{E1+E2}{2}}-x-p^{2}-(y+Y)^{2}/12)}\int 
\!d{\frac{(y-Y)}{2}}~e^{i{\frac{(y-Y)}{2}}(E1-E2)}  \nonumber \\
&=&\frac{1}{(2\pi)}  \delta (E1-E2)~f_{\frac{E1+E2}{2}}(x+p^{2}),
\end{eqnarray}
by virtue of the direct definition (\ref{Herero}).    

\section{Darboux Construction of Wigner Function Recursions} 
Analogous ladder operators for eigenstates corresponding to ``essentially
isospectral" pairs of partner potentials \cite{darboux} (familiar from
supersymmetric quantum mechanics) can also be defined mutatis-mutandis for 
Wigner functions and $\star$-products. They faithfully parallel the 
differential equation structures. 

Consider a positive semidefinite Hamiltonian
\begin{equation}    
H=p^2/2m + V(x) .
\end{equation}
This can be written as a $\star$-product of two operators, 
\begin{equation}    
H = Q^* \star Q = \left( {p\over\sqrt{2m}} +iW(x)\right)\star
\left( {p\over\sqrt{2m}} -iW(x)\right) ~, 
\end{equation}
provided 
\begin{equation}    
W^2- {\hbar \over\sqrt{2m}} \partial_x W = V(x)~. \label{iyakurandye} 
\end{equation}
This Riccati equation, familiar from ssQM, can be Darboux-transformed 
by changing variable for the ``superpotential" $W(x)$:  
\begin{equation}    
W =-\frac{\hbar ~\partial_x \psi_0}{ \sqrt{2m} ~\psi_0},
\label{ondyoze} 
\end{equation}
which reduces the condition to the Schr\"odinger equation for zero eigenvalue:
\begin{equation}    
-{\hbar^2\over 2m}  \partial^2_x \psi_0 +V(x)\psi_0=0~.
\end{equation}
Also note $Q\star f_0 =0$ for the corresponding Wigner function.
It is easy to generalize this by adding a constant to $H$ to shift 
the ground state eigenvalue from zero. 

By virtue of associativity, it is evident that the partner Hamiltonian
\begin{equation}    
H'= Q\star Q^*   =  H+ {2\hbar \over\sqrt{2m}}   \partial_x W ,
\end{equation}
i.e. the one with a partner potential 
\begin{equation}    
V'= W^2 +  {\hbar \over\sqrt{2m}} \partial_x W~,\label{Omumborombanga} 
\end{equation}
has Wigner function $\star$-genstates of the same energy as those of $H$. 
Specifically,
\begin{equation}    
H\star f=  Q^*\star Q  \star f = f \star Q^*\star Q  =Ef  
\end{equation}
implies that the real functions $Q \star f \star Q^*$ are 
$\star$-genfunctions of $H'$ with the same eigenvalue $E$:
\begin{equation}    
H'\star (Q \star f \star Q^*) =  Q \star Q^* \star Q  \star f\star Q^* 
=E (Q\star f\star Q^* )~, 
\end{equation}
{\em unless} $f$ is the Wigner function corresponding to $\psi_0$,
since $Q\star f_0=0$. 

In consequence, $E'_n= E_{n+1}$ for $n\geq 0$.
Conversely, for $g$  $\star$-genfunctions of $H'$,
$Q^* \star g\star Q$ are $\star$-genfunctions of $H$ with the same eigenvalues.

Moreover, $\psi_0'\equiv 1/\psi_0$ will be an invalid zero mode 
eigenfunction of ${\bf H'}$, as seen from the sign flip in (\ref{ondyoze}) 
and (\ref{Omumborombanga}). Consequently, an unnormalized, runaway zero
energy solution of the Schr\"{o}dinger equation with $V'(x)$ will invert 
to the legitimate ground state of ${\bf H}$ and will permit construction 
of $V$ given $V'$.    

For example, starting from the trivial potential with a continuous 
(unnormalizable) spectrum,
\begin{equation}    
V'=1,
\end{equation}
and the solution
\begin{equation}    
\psi'_0=\cosh({\sqrt{2m}x\over\hbar} ),
\qquad \Longrightarrow \qquad 
W=\tanh({\sqrt{2m}x\over\hbar} ) ,
\end{equation}
results via (\ref{iyakurandye}) in the symmetric, 
reflectionless P\"{o}schl-Teller potential \cite{teller},
$V=1-2/\cosh^2 ({\sqrt{2m}x\over\hbar} )$. 
Conversely, starting from this potential, 
\begin{equation}    
V(x) = 1- \frac{2}{\cosh^2({\sqrt{2m}x\over\hbar})}  ,
\end{equation}
there is a single bound state (normalizable to $\int \psi_0^2 =2$), 
\begin{equation}    
\psi_0 =\hbox{sech}({\sqrt{2m}x\over\hbar}) , \qquad\Longrightarrow \qquad 
W=\tanh({\sqrt{2m}x\over\hbar} ) ,
\end{equation} 
so that 
\begin{equation}    
V'=1 .
\end{equation}

Thus, the Wigner function ground state (for $m=1/2$) is 
\begin{equation}    
f_0(x,p)={1\over 2\pi} \int\! \! dy ~ { e^{-iyp} \over 
2 \cosh(x/\hbar - y / 2 )\cosh(x/\hbar + y/2)   }= 
{1\over \pi}\!  \int^{\infty}_0\! \!  \!  \!  dy ~ {\cos(yp) \over 
\cosh(2x/\hbar )+\cosh(y) }  
\end{equation}
$$
= {\sin(2xp / \hbar) \over \sinh(2x/\hbar )~ \sinh(\pi p) }~.
\qquad\qquad\qquad\qquad\qquad\qquad\qquad
\qquad %%%checks with G&R 3.983.2 
$$  
(N.B.~Not positive definite, nor a function of just $H(x,p)$.)    
It may be verified directly that 
\begin{equation}    
Q\star f_0=\left( p-{i\hbar\over 2}  \partial_x - 
i \tanh({x\over \hbar}+{i\over 2} \partial_p )  \right)  f_0(x,p)=0.
\end{equation}    

This appendage of bound states to a potential generalizes \cite{kwong} 
to the hierarchy associated with the KdV equation. Specifically, 
\begin{equation}    
W(n) = n \tanh({\sqrt{2m}x\over\hbar} )
\end{equation} 
connects the reflectionless P\"{o}schl-Teller potential
\begin{equation}    
V'(x) = n^2-{n(n-1)}/{\cosh^2({\sqrt{2m}x\over\hbar})}\qquad \hbox{ to 
its contiguous}\qquad  V(x) = n^2-{n(n+1)}/{\cosh^2({\sqrt{2m}x\over\hbar})},
\end{equation}
which has one more bound state (shape-invariance). Recursively then, one may 
go in $N$ steps, with the suitable shifts of the potential by 
$2n-1$ in each step, from the constant potential to 
\begin{equation}    
V(N; x) = N^2-{N(N+1)}/{\cosh^2({\sqrt{2m}x\over\hbar})}.
\end{equation}

Shifting this potential down by $N^2$ assigns 
the energy $E=-N^2$ to the corresponding ground state 
$\psi_0(N) = {\hbox{sech}}^N(x)$ (unnormalized), which is the null-state of 
${\hbar\over \sqrt{2m}} \partial_x+W(N)$. The corresponding (unnormalized) 
Wigner function is the $\star$-null state of $Q(N)$, 
\begin{equation}    
f_0(N;x,p)=
{1\over \pi}\!  \int^{\infty}_0\! \!  \!  \!  dy ~ {\cos(yp) \over 
(\cosh(2x/\hbar )+\cosh(y))^N }  
={1\over (N-1)!}~ 
\left(\frac{-\hbar}{2 \sinh(2x/\hbar )}  \partial_x\right)^{N-1} f_0(1;x,p),
\end{equation}
where the integral only need be evaluated from the above $f_0(1;x,p)$. 
Alternatively, 
\begin{equation}    
f_0(N;x,p)=
\left(\hbox{sech} (x/\hbar)\star\right)^{N-1} f_0(1;x,p)
\left(\star \hbox{sech} (x/\hbar)\right)^{N-1} ~.
\end{equation}

The (unnormalized) state above the ground state at $E=-(N-1)^2$ is 
$\left( {\hbar\over \sqrt{2m}} \partial_x - W(N) \right) \psi_0(N-1)$, and its 
corresponding Wigner function (setting $m=1/2$) is found recursively 
from the ground state of $H(N-1)$, through $Q^*(N)\star f_0(N-1) \star Q(N)$,
\begin{equation}    
\left( p \star f_0(N-1)+ iN\tanh({x\over\hbar} )\star f_0(N-1)\right )
 \star Q(N)=\left( p \star f_0(N-1)+ 
{N\over N-1} p \star f_0(N-1) \right)  \star Q(N) 
\end{equation}
$$
=  ({2N-1\over N-1})^2 ~p \star f_0(N-1) \star p~,
$$
by virtue of 
\begin{equation}    
Q(N-1)\star f_0(N-1)=0=f_0(N-1) \star Q^* (N-1).
\end{equation}

The state above that, at $E=-(N-2)^2$, is found recursively through
\begin{equation}
Q^*(N) \star Q^*(N-1) \star f_0(N-2)\star Q(N-1)\star Q(N)~, 
\end{equation}
and so forth. 
Thus, the entire Wigner $\star$-genfunction spectrum of $H(N)$ is obtained
with hardly any reliance on Schr\"{o}dinger eigenfunctions.

\section {Canonical Transformation of the Wigner Function}
For notational simplicity, take $\hbar=1$ in this section.
 The area element in phase space is preserved by Canonical Transformations 
\begin{equation}
(x,p) \mapsto (X(x,p), P(x,p)) 
\end{equation}
which yield trivial Jacobians ($dXdP= dx dp ~\{  X, P  \}$) by  
preserving Poisson Brackets 
\begin{equation} 
\{u,v \}_{xp}\equiv {\partial   u \over \partial x  }
{\partial v    \over \partial p }-
{\partial u    \over \partial p  }
{\partial v   \over \partial x }.
\end{equation}
They thus preserve the ``canonical invariants" of their functions:
\begin{equation}
\{  X, P  \}_{xp} = 1  \qquad\qquad  \hbox{hence} \qquad\qquad
\{  x, p  \}_{XP} = 1. 
 \end{equation}
Equivalently, 
\begin{equation}
\{  x, p  \}=\{  X, P  \},
 \end{equation}
in any basis. Motion being a canonical transformation, Hamilton's 
classical equations of motion are preserved, 
for ${\cal H}(X,P)\equiv H(x,p)$, as well \cite{uematsu}. 
What happens upon quantization?

Since, in deformation quantization, the Hamiltonian is a c-number 
function, and so transforms ``classically", ${\cal H}(X,P)\equiv H(x,p)$, 
the effects of a canonical transformation on the quantum $\star$-genvalue 
equation of Lemma 1 will be carried by a suitably transformed Wigner function. 
Predictably, the answer can be deduced from Dirac's quantum transformation 
theory. Consider the canonical transformations generated by $F(x,X)$:
\begin{equation}
p={\partial F(x,X) \over \partial x } , \qquad \qquad \qquad 
P=-{\partial F(x,X) \over \partial X }.
 \end{equation}
Following Dirac's celebrated exponentiation \cite{pamd} of such a 
generator, in the implementation of \cite{curtghand,pcm}, the energy 
eigenfunctions transform canonically 
through a generalization of the ``representation-changing" Fourier 
transform. Namely, 
\begin{equation}
\psi _{E}(x)=N_{E}\int dX\,e^{iF(x,X)}\,\Psi _{E}(X) ~.
\end{equation}
Thus, 
\begin{equation}
f(x,p)={\frac{\left| N_{E}\right| }{2\pi }}^{2}\int \!dy~\int
dX_{1}\,e^{-iF^{*}(x-y/2,X_{1})}\,\Psi _{E}^{*}(X_{1})\,e^{-iyp}\,\int
dX_{2}\,e^{iF(x+y/2,X_{2})}\,\Psi _{E}(X_{2})\;.\label{okanumaihi}
\end{equation}

The pair of Wigner functions in the respective canonical variables, 
$f(x,p)$ and 
\begin{equation} 
{\cal F}(X,P)={1\over 2\pi} \int\! dY ~\Psi^* (X-{\hbar\over2} Y )~e^{-iYP} 
\Psi(X+{\hbar\over2} Y),  
\end{equation}
are connected by a transformation functional 
${\hbox{\gothic T}}(x,p;X,P)$, 
\begin{equation}    
f(x,p)=\int \!dX\int \!dP\;\hbox{\gothic T} (x,p;X,P)
\;\bigcirc\kern-0.9em\star\; {\cal F}(X,P)=
\int \!dX\int \!dP\;\hbox{\gothic T} (x,p;X,P)
\; {\cal F}(X,P)\;, \label{Mukuru} 
\end{equation}
where $\;\bigcirc \kern-0.8em\star \;$ is with respect to the variables $X$
and $P$. 

To find this functional, let $X=\frac{1}{2}(X_{1}+X_{2})$ and 
$Y=X_{2}-X_{1}$, so that 
$\int dX_{1}\,\int dX_{2} =\int dX\,\int dY$. 
Noting that 
\begin{equation} 
\Psi^* (X-{\hbar\over2} Y )~\Psi(X+{\hbar\over2} Y)
=\int\! dP ~e^{iYP} {\cal F}(X,P)~,  
\end{equation}
it follows that (\ref{okanumaihi}) reduces to 
\begin{eqnarray}
f(x,p)&=&\frac{|N|^2}{2\pi} \int \!\!dy\int\! dX_{1}
\,e^{-iF^{*}(x-y/2,X_{1})}\,\Psi^{*}(X_{1})\,
e^{-iyp}\int\! dX_{2}\,e^{iF(x+y/2,X_{2})}\,\Psi (X_{2})
\\
&=&\frac{|N|^2}{2\pi} \int\! \!dX dY dy\,e^{-iyp } 
\,e^{-iF^{*}(x-y/2,X-Y/2)}\,\Psi^{*}(X-Y/2) \,
\Psi (X+Y/2)\,e^{iF(x+y/2,X+Y/2)} \nonumber \\
&=&\frac{|N|^2}{2\pi} \int\! \!dX dP dY dy ~ 
e^{-iyp+iPY-iF^{*}(x-y/2 ,X-Y/2 ) 
+iF(x+y/2,X+Y/2)} ~{\cal F}(X,P)~,  
\nonumber
\end{eqnarray}   
which leads to 

\noindent {\bf Lemma 3}: ${\hbox{\gothic T}}(x,p;X,P)
=\frac{|N|^2}{2\pi} \!\int\!dY dy~ \exp \left ( -iyp+iPY 
-iF^{*}(x-{\frac{y}{2}},X-{\frac{Y}{2}}) 
+iF(x+{\frac{y}{2}} ,X+{\frac{Y}{2}})\right )$. 
{\rule{.4\baselineskip}{.44\baselineskip}}

\noindent {\bf Corollary 2}: {\em This phase-space transformation functional 
obeys the ``two-star" equation,} 
\begin{equation}
H(x,p)\star {\hbox{\gothic T}}(x,p;X,P)
={\hbox{\gothic T}}(x,p;X,P)\;\bigcirc \kern-0.9em\star ~
{\cal H}(X,P)\;,\label{omunene}
\end{equation}
as follows from $H(x,-i\partial_x)\exp(iF(x,X)) ={\cal H}(X,i\partial_X)
\exp(iF(x,X))$.   
If ${\cal F}$ satisfies a $\;\bigcirc \kern-0.9em\star-$genvalue
equation, then $f$ satisfies a $\star $-genvalue equation with 
the same eigenvalue, and vice versa. 
{\rule{.4\baselineskip}{.44\baselineskip}}

Note that, by virtue of the 
spectral projection feature (\ref{omuhona},\ref{Nguarorerue}), this 
equation is also solved by any representation-changing 
equal-energy bilinear in real Wigner $\star$-genfunctions of $H$ and 
${\cal H}$,
\begin{equation}
{\hbox{\gothic T}}(x,p;X,P)=\sum_{E}g(E)~f_{E}(x,p)~{\cal F}_{E}(X,P)\;,
\end{equation}
for arbitrary real $g(E)$. Such a bilinear transformation functional is
nonsingular (invertible) if and only if $g(E)$ has no zeros on the spectrum of
either Hamiltonian\footnote{
In general, if the transformation functional effects a map to a free
particle, the $P$ integration is trivial in $(\ref{Mukuru})$, 
and the result for the Wigner function of the $x,p$ theory is just an
average over $X$ of the transformation functional. That is, if 
${\cal F}(X,P)=\delta (P-k(E))$, where $k(E)$ is the 
momentum-energy relation for the
free particle theory in question, 
$$
f(x,p)=\int \!dX\int \!dP\;\hbox{\gothic T}(x,p;X,P)\;{\cal F}(X,P)=\int
\!dX\;\hbox{\gothic T}(x,p;X,k(E)).
$$ 
One might then be tempted to wonder if just  
${\hbox{\gothic T}}(x,p;X,P)= \psi_{P}^*(x-\hbar X/2) ~e^{-iXp}
\psi_{P}(x+\hbar X/2)/2\pi \equiv {\hbox{\gothic G}}(x,p;X,P)$. 
However, what determines the allowed range for $P$?
It is always possible to embed any real energy spectrum into the
real line, but knowing this does not help at all to determine what
points are to be embedded. From the point of view of this paper, even 
when the spectrum is obvious, such a choice for the
transformation functional in general does not satisfy the two-$\star$ 
equation (\ref{omunene}). Rather, the equation fails by total derivatives that 
vary contingent on particularities of the case. E.g., for free-particle 
plane waves, $\psi_{E}(x)=\exp (iEx),$ so that 
$p\star{\hbox{\gothic G}}-{\hbox{\gothic G}}\bigcirc \kern-0.9em\star~P
=\partial_{X} {\hbox{\gothic G}}$. This  choice for ${\hbox{\gothic T}}$, 
then, does not yield useful information on the Wigner functions. 
}. %ENDFOOTNOTE

As an example, consider the linear potential again, which transforms to
a free particle (${\cal H}=P$) through
\begin{equation}
F=-\frac{1}{3}X^{3}-xX\qquad \Longrightarrow \quad p=-X,
\quad  x=P-X^{2}~.
\end{equation}
By direct computation,
\begin{equation}
{\hbox{\gothic T}}(x,p;X,P)=2^{2/3} Ai(2^{2/3}(x+X^{2}-P))~\delta
(p+X)=(2\pi)^2\int dE~f_{E}(x,p){\cal F}_{E}(X,P)~\delta (p+X)~.
\label{Otyikondo} 
\end{equation}
Note $N_{E}=1/\sqrt{2\pi}$ for the free particle energy eigenfunction 
normalization choice $\Psi _{E}(X)=(2\pi )^{-1/2}\exp (iEX)$. 
Thus, indeed, the free particle Wigner function, 
${\cal F}_{E}(X,P)=\delta (E-P)/(2\pi)$, transforms to
\begin{equation}
f(x,p)=\frac{1}{2\pi} \int \!dPdX~\hbox{\gothic T}~\delta (E-P) 
=  \frac{2^{2/3}}{2\pi} Ai\left(2^{2/3}(x+p^{2}-E)\right) ~,
\end{equation}
as it should; and (\ref{omunene}) is seen to be satisfied directly, by
virtue of the linearity of the respective Hamiltonians in the variables
$P,x$, conjugate to those of the arguments of $\delta (p+X)$.

The structure of the result in $\left(\ref{Otyikondo}\right) $
underscores that the linear potential is as ``close to classical'' as
one can get, in simple quantum mechanics. It has been noted before 
\cite{curtghand} that the transformation functional for linear potential 
wave functions is {\em exactly} the exponential of the classical generating
function for the canonical transformation to a free particle, and that
this is not the case for any other potential. The present result for the
transformation functional for Wigner functions is further evidence for
this ``close to classical'' behavior. The delta function $\delta (p+X)$ 
in $\left( \ref{Otyikondo}\right)$ is {\em half} of the classical story.
Were the Airy function also a delta function of its argument, we would
have an exact implementation of the $X,\;P\mapsto x,\;p$ classical
correspondence. As it is, there is some typically quantum mechanical
spread around the classical constraint $x+X^{2}-P=0$, in the form of
oscillations of the Airy function, and, in consequence, the Wigner 
functions of the free particle do not retain their delta-function 
form under the canonical transformation
to the linear potential Wigner functions. Reinstating $\hbar$
into (\ref{Herero})\footnote{ The exponent of the integrand turns to 
$~iy(E-x-p^2-\hbar^2 y^2/12)$. }, and taking the limit 
$\hbar \rightarrow 0$ converts the Airy 
function to a delta function, $\delta (x+X^{2}-P)$, thereupon producing the
complete classical correspondence between the two sets of phase space
variables, in that limit.

As already seen, there is substantial non-uniqueness in 
the choice of transformation functional. For example, for 
the linear potential again, (\ref{omunene}), 
\begin{equation}
(x+p^{2})\star {\hbox{\gothic S}}(x,p;X,P))=
{\hbox{\gothic S}} (x,p;X,P))\;\bigcirc \kern-0.9em\star\; P   
\end{equation}
is also satisfied by a different (and somewhat simpler) choice:
\begin{equation}
{\hbox{\gothic S}}(x,p;X,P)=
\exp  (-i(\frac{2}{3}\,X^{3}+2(x+p^{2}-P)X))\;.
\end{equation}
This transformation functional also converts
the free particle Wigner function, ${\cal F}_{E}(X,P)=\delta(E-P)/2\pi$, 
into an Airy function (as above) after integrating over the free
particle phase space, $\int \!dX \!dP $. 

Actually, it is not necessary to integrate over the phase space. 
In general, $\star $-multiplying a delta function spreads it out,
and yields a Fourier transform with respect to the conjugate variable. 
Thus, for the example considered,
\begin{eqnarray}
e^{ i( -\frac{2}{3}\,X^{3}-2(x+p^{2}-P)X)} \star \delta (P-E)
&=&\,e^{2iX(P-E)}\,\frac{1}{\pi }\int dZ\,e^{-2iZ(P-E)} e ^{i\left(
-\frac{2}{3}\,Z^{3}-2(x+p^{2}-P)Z\right)}  \nonumber \\
=\,e^{2iX(P-E)}\,\frac{1}{\pi }\int dZ\,e^{i\left(-\frac{2}{3}\,
Z^{3}-2(x+p^{2}-E)Z\right) } 
&=&e^{2iX(P-E)}\,2^{2/3}\,Ai(2^{2/3}(x+p^{2}-E))\;.
\end{eqnarray}
Hence, 
\begin{equation}
\int \!dX\int \!dP\;e^{i(-\frac{2}{3}\,X^{3}-2(x+p^{2}-P)X)}
\star \delta (P-E)=\,2^{2/3}\pi \,\,Ai(2^{2/3}(x+p^{2}-E))\;.
\end{equation}
Compare this to the action of the above ${\hbox{\gothic T}}(x,p;X,P)$, 
\begin{eqnarray}
\left( Ai(2^{2/3}(x+X^{2}-P))\delta (p+X)\right) \star \delta (P-E)
&=&  \\
e^{2iX(P-E)} \frac{1}{\pi }\int \!\!dZ \,
e^{-2iZ(P-E)}Ai(2^{2/3}(x+Z^{2}-P))\delta (p+Z)
&=&e^{2i(p+X)(P-E)}\,\frac{1}{\pi }\,Ai(2^{2/3}(x+p^{2}-P))\;.\nonumber 
\end{eqnarray}
Aside from innocuous normalizations, the difference in the two
transformation functionals acting on the free particle Wigner function
is just the phase factor $e^{2ip(P-E)}$, and the argument of the Airy
function, where $E$ has been replaced by $P$. Indeed, the phase factor 
precisely compensates for the different energy-eigenvalue occurring in the
argument of $Ai$, when acted upon by $(x+p^{2})\star $. Such simple phase 
factors may be used to shift a $\star $-genvalue whenever the Hamiltonian 
is linear in any variable.

\section{Illustrations using Liouville Quantum Mechanics}

Summary illustration of all the above, in particular the canonical
transformation effects on Wigner functions, is provided by the
Liouville model \cite{jackiw}. Our conventions for the model 
(which are essentially those of \cite{braaten}, with their 
$m\equiv 1/(4\pi )$ and their $g\equiv 1$) are given by
\begin{equation}
H_{Liouville}=p^{2}+e^{2x}\;.
\end{equation}
The energy eigenfunctions are then solutions of
\begin{equation}
\left( -\frac{d^{2}}{dx^{2}}+e^{2x}\right) \psi _{E}(x)=E\,\psi
_{E}(x)\;.
\end{equation}
The solutions are 
Kelvin (modified Bessel) $K$ functions, for $0<E<\infty $,
\begin{equation}
\psi_{E}(x)=\frac{1}{\pi }\,\sqrt{\sinh \left(\pi\sqrt{E}\right)}\,
K_{i\sqrt{E}}(e^{x})\;,
\end{equation}
which are normalized such that $\int_{-\infty }^{+\infty }dx\,\psi
_{E_{1}}^{*}(x)\,\psi _{E_{2}}(x)=\delta (E_{1}-E_{2})$.
There is no solution \cite{jackiw} for $E=0$. 

For completeness, consider the Fourier transform (including a 
convergence factor, necessary for $x\rightarrow -\infty $ 
to control plane wave behavior, but not for $x\rightarrow \infty $):
\begin{eqnarray}
\Phi _{E}(p+i\epsilon ) &=&\int_{-\infty }^{+\infty }dx\,e^{-ix\left(
p+i\epsilon \right) }\,\psi _{E}(x) \label{FourierK} \\
&=&\frac{1}{4\pi }\,\sqrt{\sinh (\pi \sqrt{E})}\;2^{-i\left( p+i\epsilon
\right) }\,\Gamma \left( \frac{-i\left( p+i\epsilon \right)
+i\sqrt{E}}{2}\right) \,\Gamma \left( \frac{-i\left( p
+i\epsilon \right)-i\sqrt{E}}{2}\right) \;.  \nonumber 
\end{eqnarray}
This follows, e.g., from a result in \cite{erdelyi}, Vol II, p 51, eqn (27):
\begin{equation}
\int_{0}^{+\infty }dz\;z^{\mu }K_{\nu }(z)=2^{\mu -1}\,\Gamma \left(
\frac{%
1+\mu +\nu }{2}\right) \,\Gamma \left( \frac{1+\mu -\nu }{2}\right) \;,
\end{equation}
valid for $\Re \left( 1+\mu \pm \nu \right) >0$ (i.e. the previous
transform is valid for $\epsilon >0$). The right hand side of this last 
relation clearly displays the symmetry $\nu \rightarrow -\nu $, 
which just amounts to the physical
statement that the energy eigenfunctions are non-degenerate for the
transmissionless exponential potential of the Liouville model.

Further note the effect on $\Phi _{E}(p+i\epsilon )$ of shifting $%
p\rightarrow p+2i$, using $\Gamma (1+z)=z\,\Gamma (z)$,
\begin{eqnarray}
\Phi _{E}(p+2i+i\epsilon ) &=&4\left( \frac{-i\left( p+i\epsilon 
\right)+i\sqrt{E}}{2}\right) 
\left( \frac{-i\left( p+i\epsilon \right)-i\sqrt{E}}{2}%
\right) \Phi _{E}(p+i\epsilon )  \nonumber \\
&=&\left( E-(p+i\epsilon )^{2}\right) \,\Phi _{E}(p+i\epsilon )\;.
\end{eqnarray}
So, as  $\epsilon \rightarrow 0$, 
$~\Phi _{E}(p+2i)=\left( E-p^{2}\right)\,\Phi _{E}(p)$. 
But this simple difference equation is just the Liouville
energy eigenvalue equation in the momentum basis,
\begin{equation}
\left( p^{2}-E\right) \Phi _{E}(p)+e^{2i\partial _{p}}\,
\Phi _{E}(p)=0\;.
\end{equation}
Such first order difference equations invariably lead to gamma
functions \cite{bender}. Below, it turns out that the Wigner 
functions also satisfy momentum difference equations, but of second order.

Many, if not all, properties of the Liouville wave functions may be
understood from the following integral representation \cite{watson},  
Ch VI, \S 6.22, eqn (10). Explicitly emphasizing the abovementioned
non-degeneracy, 
\begin{equation}
K_{ik}(e^{x})=K_{-ik}(e^{x})=\frac{1}{2}\,e^{\pi k/2}\,\int_{-\infty
}^{+\infty }dX\,e^{ie^{x}\sinh X}\,e^{ikX}.  \label{Watson}
\end{equation}
(Also see \cite{abramowitz}, 9.6.22.) This integral representation may 
be effectively regarded as the canonical transformation of a free 
particle energy eigenfunction, $e^{ikX}$, through use of the 
generating function $F(x,X)=$ $e^{x}\sinh X$. Classically, 
$p=\partial F/\partial x=e^{x}\sinh X$, and 
$P=-\partial F/\partial X=-e^{x}\cosh X$, so $P^{2}-p^{2}=e^{2x}$. 
That is, $H_{Liouville}={\cal H}_{free} \equiv P^{2}$ 
under the classical effects of the canonical transformation. 
The quantum effects are detailed below, by $\star $-acting 
with the Liouville and free Hamiltonians on the 
suitable transformation functional.

The Liouville Wigner function may be obtained from the definition 
$\left( \ref{orururumo}\right) $ in terms of known higher
transcendental functions,
\begin{eqnarray}
&&f(x,p)=\frac{1}{2\pi }\int_{-\infty }^{+\infty }dy\,
\frac{1}{\pi^{2}} \,\sinh (\pi
\sqrt{E})\,K_{i\sqrt{E}}(e^{x-y/2})\,e^{-iyp}\,
K_{i\sqrt{E}}(e^{x+y/2})\; \label{LiouvilleWigner} \\
&=&\frac{1}{4\pi^{3}}\sinh (\pi\sqrt{E})2^{2ip}
e^{(-1-2ip)x} G_{0\,4}^{4\,0}\left( 
\left. \frac{e^{4x}}{16}\right| \frac{1+2i\sqrt{E}}{4},
\frac{1-2i\sqrt{E}}{4},\frac{1+2i\sqrt{E}+4ip}{4},
\frac{1-2i\sqrt{E}+4ip}{4}\,\right).  \nonumber
\end{eqnarray}
The following K-transform was utilized to express this result 
in closed form,
\begin{equation}
\int_{0}^{\infty }dw\,(wz)^{1/2}\,w^{\sigma -1}\,K_{\mu }
\left( a/w\right)
\,K_{\nu }(wz)=2^{-\sigma -5/2}a^{\sigma }\,G_{04}^{40}\left( 
\left.  \frac{a^{2}z^{2}}{16}\right| \frac{\mu -\sigma }{2}
\,,\frac{-\mu -\sigma}{2},\frac{1}{4}+\frac{\nu }{2}
\,,\frac{1}{4}-\frac{\nu }{2}\right) .
\label{Meijer}
\end{equation}
The right hand side involves a special case of Meijer's G-function,
\begin{equation}
G_{pq}^{mn}\left( z\left|
\begin{array}{c}
a_{i},\;i=1,\ldots ,p \\
b_{j},\;j=1,\ldots ,q
\end{array}
\right. \right) ,
\end{equation}
(cf.~\cite{erdelyi}, \S 5.3), which is fully symmetric in the 
parameter subsets $\{a_{1},\ldots ,a_{n}\}$, 
$\{a_{n+1},\ldots,a_{p}\},\;\{b_{1},\ldots ,b_{m}\}$, 
and $\{b_{m+1},\ldots ,b_{q}\}$. It is possible to re-express the 
result as a linear combination of generalized
hypergeometric functions of type $_{0}F_{3}$, but there is little reason
to do so here. This transform is valid for $\Re a>0$, and is taken from
\cite{meijer}, p 711, eqn (55).\footnote{%
There is an error in this result as it appears in \cite{erdelyiInt}, Vol
II,  \S 10.3, eqn (58), where the formula has $a^{2}z^{2}/4$ instead
of $a^{2}z^{2}/16$ as the argument of the G-function. The latter
argument is correct, and appears in Meijer's original paper cited here.}
The transform is complementary to \cite{erdelyiInt}, \S 10.3, eqn (49), 
in an obvious way, a K-transform which appears in perturbative 
computations of certain Liouville correlation functions \cite{braaten}.

The result $\left( \ref{LiouvilleWigner}\right) $ may be written in
slightly different alternate forms,
\begin{eqnarray}
f(x,p) &=&\frac{\sinh (\pi \sqrt{E})e^{-x}}
{4\pi^{3}}G_{0\,4}^{4\,0}
\left( \left. \frac{e^{4x}}{16}\right|
\frac{1+2i\sqrt{E}-2ip}{4},
\frac{1-2i\sqrt{E}-2ip}{4},\frac{1+2i\sqrt{E}+2ip}{4},
\frac{1-2i\sqrt{E}+2ip}{4}\,\right)   \nonumber \\
&=&\frac{\sinh (\pi \sqrt{E})\;}{8\pi ^{3}}\,G_{0\,4}^{4\,0}\left(
\left.
\frac{e^{4x}}{16}\right|
\frac{i\sqrt{E}-ip}{2},\frac{-i\sqrt{E}-ip}{2},%
\frac{i\sqrt{E}+ip}{2},\frac{-i\sqrt{E}+ip}{2}\,\,\right) ,
\label{LiouvilleWignerAlt}
\end{eqnarray}
by making use of the parameter translation identity for the G-function
(\cite{erdelyi} \S 5.3.1, eqn $\left( 8\right) $):
\begin{equation}
z^{\lambda }G_{pq}^{mn}\left( z\left|
\begin{array}{l}
a_{r} \\
b_{s}
\end{array}
\right. \right) =G_{pq}^{mn}\left( z\left|
\begin{array}{l}
a_{r}+\lambda  \\
b_{s}+\lambda
\end{array}
\right. \right) .  \label{TranslIdent}
\end{equation}

Yet another way to express the result utilizes the Fourier transform of
the wave function, $\left( \ref{FourierK}\right)$, in terms of which the 
Wigner function reads, in general,  
\begin{equation}
f(x,p)=\left( {\frac{1}{2\pi }}\right) ^{2}\int_{-\infty }^{+\infty
}dk\,\Phi _{E}^{*}(p-{\frac{1}{2}}k)\,e^{ixk}\,\Phi
_{E}(p+{\frac{1}{2}}k)\;.
\end{equation}
The specific result $\left( \ref{FourierK}\right)$ then gives, as
$\epsilon\rightarrow 0$,
\begin{eqnarray}
f(x,p) &=&\left( \frac{1}{8\pi ^{2}}\right) ^{2}\,\sinh (\pi \sqrt{E}%
)\,\int_{-\infty }^{+\infty }dk\;e^{ixk}\,4^{-i\left( k/2+i\epsilon
\right)
}\,\Gamma \left( \frac{i\left( p-k/2-i\epsilon \right)
-i\sqrt{E}}{2}\right)
\times  \\
&\times& \,\Gamma \left( \frac{i\left( p-k/2-i\epsilon \right)
+i\sqrt{E}}{2}%
\right) \,\Gamma \left( \frac{-i\left( p+k/2+i\epsilon \right)
+i\sqrt{E}}{2}%
\right) \,\Gamma \left( \frac{-i\left( p+k/2+i\epsilon \right)
-i\sqrt{E}}{2}\right).  \nonumber
\end{eqnarray}
However, this is a contour integral representation of the
particular G-function given above. Because of the $\epsilon$ 
prescription, the contour
in the variable $z=k/2+i\epsilon $ runs parallel to the real axis, but
slightly above the poles of the $\Gamma $-functions located on the real
axis at $z=p-\sqrt{E},\;z=p+\sqrt{E},\;z=-p+\sqrt{E},$
and$\;z=-p-\sqrt{E}\;.$
Changing variables to $s=\frac{1}{2}iz$ yields 
\begin{eqnarray}
f(x,p) &=&\frac{1}{8\pi ^{3}}\,\sinh (\pi \sqrt{E})\;\frac{1}{2\pi i}%
\,\int_{C}ds\,\left( \frac{e^{4x}}{16}\right) ^{s}\Gamma \left(
\frac{ip-i\sqrt{E}}{2}-s\right) \times   \label{MellinBarnes} \\
&&\times \,\Gamma \left(\frac{ip+i\sqrt{E}}{2}-s\right) \,
\Gamma \left(\frac{-ip+i\sqrt{E}}{2}-s\right) \,\Gamma \left(
\frac{-ip-i\sqrt{E}}{2}-s\right) \;,  \nonumber
\end{eqnarray}
where the contour $C$ in the $s$-plane runs from $-i\infty $ to
$+i\infty $, just to the left of the four poles on the imaginary 
$s$ axis at $i(p+\sqrt{E}) /2,\;i( p-\sqrt{E})/2,\;i( -p+\sqrt{E})/2,$ 
and $\;i(-p-\sqrt{E})/2$. This is recognized as
the Mellin-Barnes type integral definition of the G$_{04}^{40}$-function
(cf.~\cite{erdelyi}, \S 5.3, eqn (1)) in agreement with the second result
above, $\left( \ref{LiouvilleWignerAlt}\right) $.

The translation identity $\left( \ref{TranslIdent}\right) $ is seen to hold 
by virtue of $\left( \ref{MellinBarnes}\right) $, through simply shifting the
variable of integration, $s$. Moreover, deforming the contour in $\left(
\ref{MellinBarnes}\right) $ to enclose the four sequences of poles 
$s_{n}=n+i( \pm p\pm \sqrt{E})/2$ reveals the equivalence of
this particular G-function to a linear combination of four $_{0}F_{3}$
functions, one for each of the sequences of poles. Evaluating the
integral by the method of residues for all these poles produces the standard
$_{0}F_{3}$ hypergeometric series.

It should now be straightforward to directly check that the explicit
result for $f(x,p)$ is indeed a solution to the Liouville 
$\star $-genvalue equation, 
\begin{equation}
H_{Liouville}\star f(x,p)=\left( (p-{\frac{i}{2}}\partial
_{x})^{2}+e^{2(x+{%
\frac{i}{2}}\partial _{p})}\right) \,f(x,p)=E\,f(x,p)\;.
\label{Liouville*Eqn}
\end{equation}
For real $E$ and real $f(x,p)$, the imaginary part of this 
$\star$-genvalue equation is
\begin{equation}
\left( -{p}\partial _{x}+e^{2x}\sin \partial _{p}\right) \,f(x,p)=0\;,
\end{equation}
while the real part is
\begin{equation}
\left( p^{2}-E-{\frac{1}{4}}\partial _{x}^{2}+e^{2x}\cos \partial
_{p}\right) \,f(x,p)=0\;.
\end{equation}
The first of these is a first-order differential/difference equation
relating the $x$ and $p$ dependence, 
\begin{equation}
\,e^{-2x}\partial _{x}f(x,p)=\frac{1}{2ip}\left(
f(x,p+i)-f(x,p-i)\right) .
\end{equation}
Similarly, the real part of the $\star $-genvalue equation is a
second-order differential/difference equation,
\begin{equation}
e^{-2x}\left( p^{2}-E-{\frac{1}{4}}\partial _{x}^{2}\right)
\,f(x,p)+\frac{1%
}{2}\left( f(x,p+i)+f(x,p-i)\right) \,=0\;.
\end{equation}
The previous first-order equation may now be substituted (twice) into
this last second-order equation, to convert it from a 
differential/difference equation into a second-order 
difference only equation in the momentum
variable, with non-constant coefficients. That is,
\begin{eqnarray}
0&=&\left( p^{2}-E\right) \,f(x,p)+\left( \frac{e^{2x}}{4p}\right)
^{2}\,\left( f(x,p+2i)-2f(x,p)+f(x,p-2i) \right)   \nonumber \\
&&+i\,\frac{e^{2x}}{4p}\left( f(x,p+i)-f(x,p-i) \right)
+\frac{e^{2x}}{2}\left( f(x,p+i)+f(x,p-i)\right) .
\end{eqnarray}
We leave it as an exercise for the reader to exploit the recursive
properties of the Meijer G-function and show that this difference
equation is indeed obeyed by the result 
$\left( \ref{LiouvilleWigner}\right)$.
Rather than pursue this in detail, we turn our attention to the
transformation functional which connects the above result for $f$ to a
free particle Wigner function.

Given $\left( \ref{Watson}\right) $, it follows that 
\begin{equation}
\psi _{E}(x)=\frac{1}{\pi }\,\sqrt{\sinh (\pi \sqrt{E})}\;K_{i\sqrt{E}%
}(e^{x})=\frac{1}{2\pi }\,\sqrt{\sinh (\pi \sqrt{E})}\;\,e^{\pi
\sqrt{E}/2}\,\int_{-\infty }^{+\infty }dX\,e^{ie^{x}\sinh X}\,
e^{i\sqrt{E}X}\;,
\end{equation}
hence 
$N_E=(4 \pi\sqrt{E}e^{\pi\sqrt{E}}\sinh(\pi\sqrt{E}))^{1/2}/{2\pi} $,
if we choose a $\delta (E_1 - E_2)$ normalization for the free particle
plane waves as well as for the Liouville eigenfunctions. Therefore, Lemma 3 
yields
\begin{equation}
{\hbox{\gothic T}}(x,p;X,P)=\frac{\left| N\right| ^{2}}{2\pi }\int
\!dYdy~\exp \left( -iyp+iPY-iF^{*}(x-y/2,X-Y/2)+iF(x+y/2,X+Y/2)\right)
\nonumber
\end{equation}
$$
=\frac{1}{\left( 2\pi \right) ^{3}}\left(4 \pi \sqrt{E}e^{\pi \sqrt{E}}
\sinh (\pi\sqrt{E})\right) \int \! \! dYdy~\exp\left( -iyp+iPY
-ie^{x-y/2}\sinh \left( X-\frac{Y}{2}\right) +ie^{x+y/2}\sinh 
\left( X+\frac{Y}{2}\right) \right)
$$
\begin{eqnarray*}
&=&\frac{1}{4\pi ^{3}}\left( 4 \pi \sqrt{E}e^{\pi \sqrt{E}}
\sinh (\pi \sqrt{E})\,\right) \int 
\,\!d\left( \frac{y+Y}{2}\right) \exp \left( i\left(
P-p\right) \frac{y+Y}{2}+ie^{x+X}\,\sinh \left( \frac{y+Y}{2}\right)
\right)
\times \qquad   \qquad   \qquad   \qquad   \\
&&\times \int d\left( \frac{Y-y}{2}\right) \,\exp \left( i\left(
P+p\right)
\frac{Y-y}{2}+ie^{x-X}\sinh \left( \frac{Y-y}{2}\right) \right) \;.
\end{eqnarray*}
We thus conclude,
\begin{equation}
{\hbox{\gothic T}}(x,p;X,P)=\frac{4} {\pi^2} \sqrt{E}
e^{\pi\sqrt{E}}\sinh (\pi \sqrt{E})\,\,e^{-\pi P}\;K_{i\left( P-p\right)
}(e^{x+X})\,K_{i\left( P+p\right) }(e^{x-X})\;.  \label{Omakungurua} 
\end{equation}

We now check that this result obeys $\left( \ref{omunene}\right) $ and,
in so doing, carry out the nontrivial steps needed to 
show the Liouville Wigner
functions satisfy the Liouville $\star $-genvalue equation $\left( \ref
{Liouville*Eqn}\right) $.
That is to say, we shall show  
\begin{equation} 
\left((p-{\frac{i}{2}}\overrightarrow{\partial }_{x})^{2}
+e^{2(x+{\frac{i}{2}}\overrightarrow{\partial }_{p})}\right) 
{\hbox{\gothic T}}(x,p;X,P)=
{\hbox{\gothic T}}(x,p;X,P)\left( (P+{\frac{i}{2}}%
\overleftarrow{\partial }_{X})^{2}\right) ~, 
\end{equation}
or, equivalently,
\begin{equation}
\left( (p-{\frac{i}{2}}\overrightarrow{\partial
}_{x})^{2}+e^{2(x+{\frac{i}{2%
}}\overrightarrow{\partial
}_{p})}-(P+{\frac{i}{2}}\overrightarrow{\partial }%
_{X})^{2}\right) \,K_{i\left( P-p\right) }(e^{x+X})\,K_{i\left(
P+p\right)
}(e^{x-X})=0\;.  \label{Oomakungurua} 
\end{equation}
Specifically, 
\begin{equation}
\frac{-1}{4}\left(\overrightarrow{\partial}_{x}^{2}-
\overrightarrow{\partial}_{X}^{2}\right) K_{i\left( P-p\right)}(e^{x+X})
\,K_{i\left(P+p\right)}(e^{x-X})=-e^{2x}\,
K_{i\left(P-p\right)}^{\prime}(e^{x+X})\,K_{i\left(P+p\right)}^{\prime}
(e^{x-X})\;,  \label{part1}
\end{equation}
\begin{eqnarray}
&&\left(-ip\overrightarrow{\partial}_x-iP\overrightarrow{\partial}_X
\right) K_{i\left( P-p\right) }(e^{x+X})\,K_{i\left( P+p\right)}
(e^{x-X})  \label{part2} \\
&&=-i\left( p+P\right) e^{x+X}\,K_{i\left( P-p\right) }^{\prime
}(e^{x+X})\,K_{i\left( P+p\right) }(e^{x-X})-i\left( p-P\right)
e^{x-X}\,K_{i\left( P-p\right) }(e^{x+X})\,K_{i\left( P+p\right)
}^{\prime}(e^{x-X})\;,  \nonumber
\end{eqnarray}
and
\begin{equation}
e^{2(x+{\frac{i}{2}}\overrightarrow{\partial }_{p})}\,K_{i\left(
P-p\right)
}(e^{x+X})\,K_{i\left( P+p\right) }(e^{x-X})=e^{2x}\,K_{1+i\left(
P-p\right)
}(e^{x+X})\,K_{-1+i\left( P+p\right) }(e^{x-X})\,\;.  \label{part3}
\end{equation}

Now, recall the recurrence relations (\cite{abramowitz}, 9.6.26)
\begin{eqnarray}
K_{1+i\left( P-p\right) }(e^{x+X}) &=&-K_{i\left( P-p\right) }
^{\prime}(e^{x+X})
+i\left( P-p\right) e^{-x-X}\,K_{i\left( P-p\right)}(e^{x+X})\;,
\\
K_{-1+i\left( P+p\right) }(e^{x-X}) &=&-K_{i\left( P+p\right) }
^{\prime}(e^{x-X})-i\left( P+p\right) e^{-x+X}\,
K_{i\left( P+p\right)}(e^{x-X})\;.
\end{eqnarray}
So the previous relation $\left( \ref{part3}\right) $ becomes
\begin{eqnarray}
e^{2(x+{\frac{i}{2}}\overrightarrow{\partial }_{p})}\,
K_{i\left(P-p\right)}(e^{x+X})\,
K_{i\left( P+p\right) }(e^{x-X}) &=&e^{2x}\,
K_{i\left(P-p\right) }^{\prime }(e^{x+X})\,
K_{i\left( P+p\right) }^{\prime}(e^{x-X})\label{part4} \\
&&+i\left( P+p\right) e^{x+X}\,K_{i\left( P-p\right)}
^{\prime}(e^{x+X})\,K_{i\left( P+p\right) }(e^{x-X})\nonumber \\
&&-i\left( P-p\right) e^{x-X}\,
K_{i\left( P-p\right)}(e^{x+X})\,K_{i\left(
P+p\right) }^{\prime }(e^{x-X})  \nonumber \\
&&+\left( P^{2}-p^{2}\right) \,
K_{i\left( P-p\right)}(e^{x+X})\,
K_{i\left(P+p\right) }(e^{x-X}).\nonumber
\end{eqnarray}
The sum of $\left( \ref{part1}\right) $, $\left( \ref{part2}\right) $, 
and  $\left( \ref{part4}\right) $, shows that 
$\left(\ref{Oomakungurua}\right)$ is, indeed, satisfied.

Integrating over $X$ and $P$ the product of ${\hbox{\gothic T}}(x,p;X,P)$ 
and the free particle Wigner function, as given here by 
$(4 \pi \sqrt{E})^{-1}\delta (P-\sqrt{E})$, 
yields another expression for the Liouville
Wigner function which checks against the previous result,
$\left(\ref{LiouvilleWigner}\right) $. 
Using $\left( \ref{Meijer}\right) $ and the
parameter translation identity for the G-function, this other expression
is just  $\left( \ref {LiouvilleWignerAlt}\right) $.

Supersymmetric Liouville quantum mechanics is obtained by carrying
through the Darboux construction detailed above (with $\hbar =1=2m$), 
for the choice
\begin{equation}
W\left( x\right) =e^{x}.
\end{equation}
Conventions essentially follow \cite{tcgg}. 

The first Hamiltonian of the essentially isospectral pair is then
\begin{equation}
H=p^{2}+e^{2x}-e^{x},
\end{equation}
and the allowed spectrum is $0\leq E<\infty \,$, 
including zero-energy, for which there is a bounded wave function 
normalized as part of the continuum, 
\begin{equation}
\psi _{0}(x)=\frac{1}{\sqrt{\pi }\,}\,e^{-e^{x}}.
\end{equation}
The other, $E>0$, eigenfunctions are
\begin{equation}
\psi_{E}(x)=\left( \frac{1}{4\pi ^{2}\sqrt{E}}\,e^{x}\cosh \left( 
\pi\sqrt{E}\right) \right) ^{1/2}\,\left( K_{\frac12
-i\sqrt{E}}(e^{x})+K_{\frac12 
+i\sqrt{E}}(e^{x})\right) \;,
\end{equation}
again normalized so that $\int_{-\infty }^{+\infty }dx\,\psi
_{E_{1}}^{*}(x)\,\psi _{E_{2}}(x)=\delta (E_{1}-E_{2})$.

The second Hamiltonian of the pair is
\begin{equation}
H^{\prime }=p^{2}+e^{2x}+e^{x},
\end{equation}
and the allowed spectrum is $0<E<\infty $, excluding 
zero energy\footnote{ 
The candidate 
$\psi _{0}^{\prime }(x)=1/\psi _{0}(x)=\sqrt{\pi }\,\exp (e^x)$ 
solves the Schr\"{o}dinger equation, but is obviously unbounded,
as expected.}. The $E>0$ eigenfunctions are then 
\begin{equation}
\psi _{E}^{\prime }(x)=\left( \frac{1}{4\pi ^{2}\sqrt{E}}\,e^{x}\cosh
\left(\pi \sqrt{E}\right) \right) ^{1/2}\,
\left( iK_{\frac12 -i\sqrt{E}}(e^{x})-iK_{\frac12 
+i\sqrt{E}}(e^{x})\right) \;,
\end{equation}
and may be obtained from the previous $E>0$ eigenfunctions, as 
$\psi_{E}^{\prime }(x)=\frac{1}{\sqrt{E}}\,
\left( \partial_{x}+W\right) \psi_{E}(x)$.

For both Hamiltonians, the Wigner functions are straightforward to
construct directly, once again leading to the K-transform 
$\left( \ref{Meijer}\right) $ and particular
Meijer G-functions. We find it sufficient here to consider only the 
ground state for $H$,
\begin{equation}
f_{0}(x,p) =\frac{1}{2\pi ^{2}}\,
\int_{-\infty }^{+\infty}dy\,e^{-2e^{x}\cosh (y/2)-iyp}  
=\frac{2}{\pi ^{2}}\,K_{2ip}(2e^{x})\;,
\end{equation} 
a single modified Bessel function. It smoothly satisfies 
$(p-iW(x))\star f_0=0$, 
and hence the $\star$-genvalue equation $H\star f_{0}=0$.

\noindent{\Large{\bf Acknowledgements}} 

We thank Y Hosotani for helpful discussions.
This work was supported in part by NSF grant PHY 9507829, and
by the US Department of Energy, Division of High Energy Physics, 
Contract W-31-109-ENG-38.
%%%%%%%   %%%%%%%   %%%%%%%   %%%%%%%   %%%%%%%   %%%%%%%   

\end{document}